# Probing the phase transition to a coherent 2D Kondo lattice

*Cosme G. Ayani[1,2], Michele Pisarra[3], Iván M. Ibarburu[1], Manuela Garnica[2,6], Rodolfo Miranda[1,2,4,6], Fabián Calleja[2,*], Fernando Martín[2,4,5], Amadeo L. Vázquez de Parga[1,2,4,6,*]*

[1]Departamento de Física de la Materia Condensada, Universidad Autónoma de Madrid, Cantoblanco 28049, Madrid, Spain.

[2]IMDEA Nanociencia, Calle Faraday 9, Cantoblanco 28049, Madrid, Spain.

[3] Dipartimento di Física, Università della Calabria, Via P. Bucci, Cubo 30C and INFN, Sezione LNF, Gruppo collegato di Cosenza, Cubo 31C, 87036 Rende (CS), Italy

[4] IFIMAC, Universidad Autónoma de Madrid, Cantoblanco 28049, Madrid, Spain

[5] Dep. Química Módulo 13, Universidad Autónoma de Madrid, Cantoblanco, 28049 Madrid, Spain.

[6] Instituto Nicolás Cabrera, Universidad Autónoma de Madrid, Cantoblanco, 28049, Madrid, Spain

*Corresponding author. Email: fabian.calleja@imdea.org, al.vazquezdeparga@uam.es

Kondo lattices are systems with unusual electronic properties that stem from strong electron correlation, typically studied in intermetallic 3D compounds containing lanthanides or actinides. Lowering the dimensionality of the system enhances the role of electron correlations providing a new tuning knob for the search of novel properties in strongly correlated quantum matter. We report the realization of a 2D Kondo lattice by stacking a single layer Mott insulator on a metallic surface. We steadily lower the temperature of the system and by using high-resolution scanning tunnelling spectroscopy we follow the phase transition leading to the Kondo lattice. Above 27 K the interaction between the Mott insulator and the metal is negligible and both keep their original electronic properties intact. Below 27 K the Kondo screening of the localized electrons in the Mott insulator begins and below 11 K the formation of a coherent quantum electronic state extended to the entire sample, i.e. the Kondo lattice, takes place. By means of density functional theory we explain the electronic properties of the system and its evolution with temperature. Our findings contribute to the exploration of unconventional states in 2D correlated materials.

## 1. Introduction

Solids containing strongly correlated electrons, whose mutual interaction energy is comparable or even larger than their kinetic energy, are the cradle of quantum states with unusual collective properties. The interplay between charge, spin and orbital degrees of freedom gives rise to unusual phenomena, as the coexistence of superconductivity and magnetism,[1,2] the appearance of multiple superconducting phases with different order parameters,[3,4] the existence of mixed valence states[5] and fractional quantum hall states,[6] or the electronic properties of quantum dots. The manipulation of electron correlations in these materials has thus direct implications in the development of superconducting magnets, magnetic storage, Mott transistors and sensing,[7,8] amid other applications.

Among existing strongly correlated materials, three-dimensional (3D) heavy fermion compounds containing 4f and 5f electrons are the ones with the strongest electron correlation in the ground state.[9] In these systems, some of the conduction electrons are highly localized inside the f-orbitals, giving rise to localized magnetic moments. The interaction between the resulting 3D lattice of magnetic moments and the conduction sea is the key ingredient for the strongly correlated physics in these systems. When the system's dimensionality is reduced from 3D to 2D, the Coulomb interaction between electrons becomes even more relevant. In addition, thermal and quantum fluctuations are largely enhanced, thus expanding the critical regions around the so-called quantum critical point.[10] Consequently, many-body effects that do not exist or manifest in 3D are expected to prevail in highly correlated 2D systems, and a paradigmatic example of these effects is the formation of two-dimensional Kondo lattices. However, although there are already several studies of 2D networks of Kondo impurities,[11-13] the experimental demonstration of the actual coherent state associated to the Kondo lattice regime in these systems has remained elusive so far.[14]

When a magnetic ion is placed in a metallic host, the screening of the magnetic moment by the conduction electrons is known as Kondo effect.[15-17] This effect manifests when conduction electrons at the Fermi level are in resonance with a flip of the two-fold degenerate spin ground state of the magnetic impurity. For an antiferromagnetic spin-exchange coupling between the localized electron of the impurity and the electrons in the conduction sea, a many-body spin-singlet state is formed below a characteristic temperature, known as Kondo temperature, $T_K$, as sketched in **Figure 1a**. The Kondo singlet formation is revealed in the Local Density of States (LDOS) as an intense zero-bias feature, the Kondo resonance, sketched as a sharp peak in **Figure 1b**. As the temperature is lowered, the conduction electrons become confined to the Fermi surface, increasing the number of electrons contributing to the resonant quantum spin-flip scattering, and the intensity of the Kondo resonance rapidly increases upon cooling to temperatures below the Kondo temperature. Figure 1b shows schematically the Kondo resonance together with the singly occupied electronic state of the impurity, $\varepsilon_d$, and the energy needed to bring an additional electron from an occupied state of the metal at the Fermi energy $E_F$ into the impurity electronic state, $\varepsilon_d+U$. The singlet bound state implies that the spin of the impurity is effectively quenched, so the remaining conduction electrons can only experience scattering by the electrostatic potential, which leads to an increase in the resistance as the system is cooled below the Kondo temperature. This increase in the resistance at low temperatures was observed in metallic wires and was the first experimental observation related to Kondo screening.[18]

The presence of several localized magnetic moments (as opposed to just one) in a metallic host can further modify the physics of the system. The same spin-exchange coupling that induces the Kondo effect can also induce a magnetic interaction between the localized spins. Indeed, the localized impurities can exchange their spins through scattering of two conduction electrons traveling between the impurity sites. This process, which is the result of the so-called Ruderman-Kittel-Kasuya-Yosida (RKKY) interaction,[19-21] favors long-range magnetic order depending on the density of states at the Fermi level, the periodicity and the distance between the magnetic moments of the impurities. Therefore, in the presence of a dense lattice of localized magnetic moments, the resulting ground state can be either magnetically ordered or a paramagnetic Fermi liquid, depending on the relative strength of the Kondo coupling and the RKKY interaction, as discussed by Doniach.[22]

Theoretical predictions of the possible existence of Kondo lattices and the physics behind this phenomenon date from as early as the 1980s.[23,24] In Kondo lattices, at temperatures below a characteristic value, $T_{KL}$, the Kondo clouds of the individual impurities are coherently superimposed and acquire the periodicity of the crystal, as sketched in **Figure 1c**. Bloch's theorem ensures the formation of a renormalized flat band of width of the order of $T_{KL}$. The electrons originally localized at the magnetic impurities are thus delocalized becoming part of the new Fermi surface, while at the same time the conduction electrons of the crystal acquire very large effective masses due to the hybridization of the new flat band with the existing conduction bands. The corresponding LDOS at the Fermi level is expected to exhibit a more complex gap-like structure,[23] as depicted in **Figure 1d**, in contrast with the single peak related to the single Kondo impurity regime (Figure 1b). The characteristic temperature $T_{KL}$ associated to the energy scale for the formation of the coherent Kondo lattice is expected to be lower than $T_K$ because, once the localized magnetic moments are lost, the magnetic interaction between impurities is no longer in operation. Hence, the energy gain per site due to the Kondo lattice development is smaller than the energy gain for a single Kondo resonance.[25]

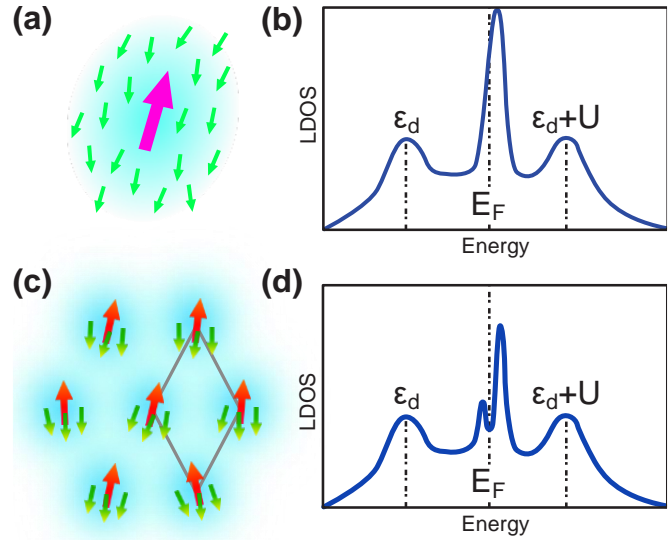

**Figure 1.** **(a)** Schematic representation of the single Kondo impurity case. The red arrow represents the spin of the impurity and the green arrows the spins of the conduction electrons **(b)** Sketch of the LDOS corresponding to the case shown in panel (a). **(c)** Schematic representation of a Kondo lattice formed from an array of impurities. Same color convention as in panel (a). **(d)** Corresponding LDOS sketch associated to panel (c). In panels (b) and (d) $\varepsilon_d$ shows the energy position of the impurity singly occupied electronic state, while $\varepsilon_d + U$ represents the energy needed for the double occupancy of that state.

The invention of the scanning tunnelling microscope (STM) has allowed for the experimental exploration of the Kondo physics on individual magnetic adatoms or charged organic molecules.[26-30] Van der Waals heterostructures offer the possibility to integrate 2D materials to produce structures with unprecedent characteristics and novel physics.[31,32] In this work we show, by means of low temperature STM and STS, the buildup of a 2D Kondo lattice in a system composed by a Mott insulator, a single 1T-$TaS_2$ layer, stacked on the surface of a metallic layered crystal, 2H-$TaS_2$ (see **Figure 2** for a schematic model). We unambiguously demonstrate the existence of the resulting collective quantum coherent phase by measuring the characteristics of the Kondo-lattice gap that develops within the Kondo resonance at the Fermi level. Upon ramping up the temperature across $T_{KL}$, we follow the evolution of this gap during the phase transition from the Kondo lattice regime to the single-Kondo impurity regime. The observed modifications in the LDOS are well explained by state-of-the-art Density Functional Theory (DFT) calculations.

TaS$_2$ is a transition metal dichalcogenide with a quasi-two-dimensional character,[33,34] which has two basic structures: a 2H-TaS$_2$ structure where the coordination between the Ta and S atoms is trigonal prismatic, and a 1T-TaS$_2$ structure with an octahedral coordination. These two polymorphs are very close in formation energy.[35] 2H-TaS$_2$ undergoes an in-plane transition around 78 K leading to a long-range quasi-commensurate (3×3) charge density wave (CDW)[36] on the sample's surface, causing the opening of a pseudo gap at the Fermi level due to the incomplete nesting in the 2D Fermi surface.[37,38] 1T-TaS$_2$ presents, below 180 K, a commensurate CDW with a periodicity of $(\sqrt{13} \times \sqrt{13})R13,9°$.[34] The atomic structure consists in clusters of 13 Ta atoms in which the 12 outer atoms are slightly displaced towards the central one, forming a Star-of-David (SoD) cluster.[34] The 5d orbitals of the 12 outer Ta atoms form the valence and conduction bands separated by the CDW gap.[39,40] The remaining 5d orbital, corresponding to the Ta atom at the center of the SoD, forms a half-filled band, suggesting a metallic character of the 1T-TaS$_2$ structure in the ground state.[41] However, experimentally, this structure is found to be insulating,[39,42] which has been attributed to a Mott localization of the remaining single electron at the center of the SoD.[43] The Mott localization splits the band at the Fermi level in two Hubbard sub-bands separated by a Mott-Hubbard gap.[41]

## 2. Results

A 2H-TaS$_2$ single crystal was cleaved and inserted in the low temperature STM without breaking the Ultra-High Vacuum (UHV) conditions (see section 1 in the Supporting Information (SI)). The sample presents terraces of the order of 2 µm wide separated by single layer steps, as can be seen in the corresponding large scale STM images (see SI section 3). Most of the surface presents a quasi-commensurate (3×3) CDW as expected for a 2H-TaS$_2$ single crystal.[36] Due to the close values of the formation energy for the 2H-TaS$_2$ and the 1T-TaS$_2$ polymorphs,[35] it was possible to find, in the 2H-TaS$_2$ crystal, areas where the topmost layer exhibited a 1T-TaS$_2$ structure. **Figure 2a** shows a representative STM image with a single layer step-edge separating two atomic terraces. The lower terrace (right-hand side) presents the quasi-(3×3) CDW periodicity expected for a 2H-TaS$_2$ crystal. An atomically resolved STM image measured on this terrace can be seen in the inset on the right. Both the atomic (green rhombus) and the CDW (blue rhombus) periodicities are indicated. The upper (left-hand side) terrace presents a different long-range periodicity, namely $(\sqrt{13} \times \sqrt{13})R13,9°$, which corresponds to the CDW present at low temperature on 1T-TaS$_2$. The left inset in Figure 2a shows an atomically resolved image measured on the upper terrace. The structure of the SoD unit cell is superimposed to the topographic image on its lower right corner. **Figure 2b** shows a line profile measured across the step, with an apparent step height of $619 \pm 50$ pm, corresponding to a single layer of TaS$_2$. The corrugation associated with the $(\sqrt{13} \times \sqrt{13})R13,9°$ periodic structure can be seen in the profile, while that of the quasi-(3×3) periodicity is not visible due to its intrinsic low magnitude in the STM images. The discrimination of the other possible 1T/1T scenarios and the determination of the periodicity of both CDWs by STM is described in detail in sections 4 and 5 of the SI, respectively.

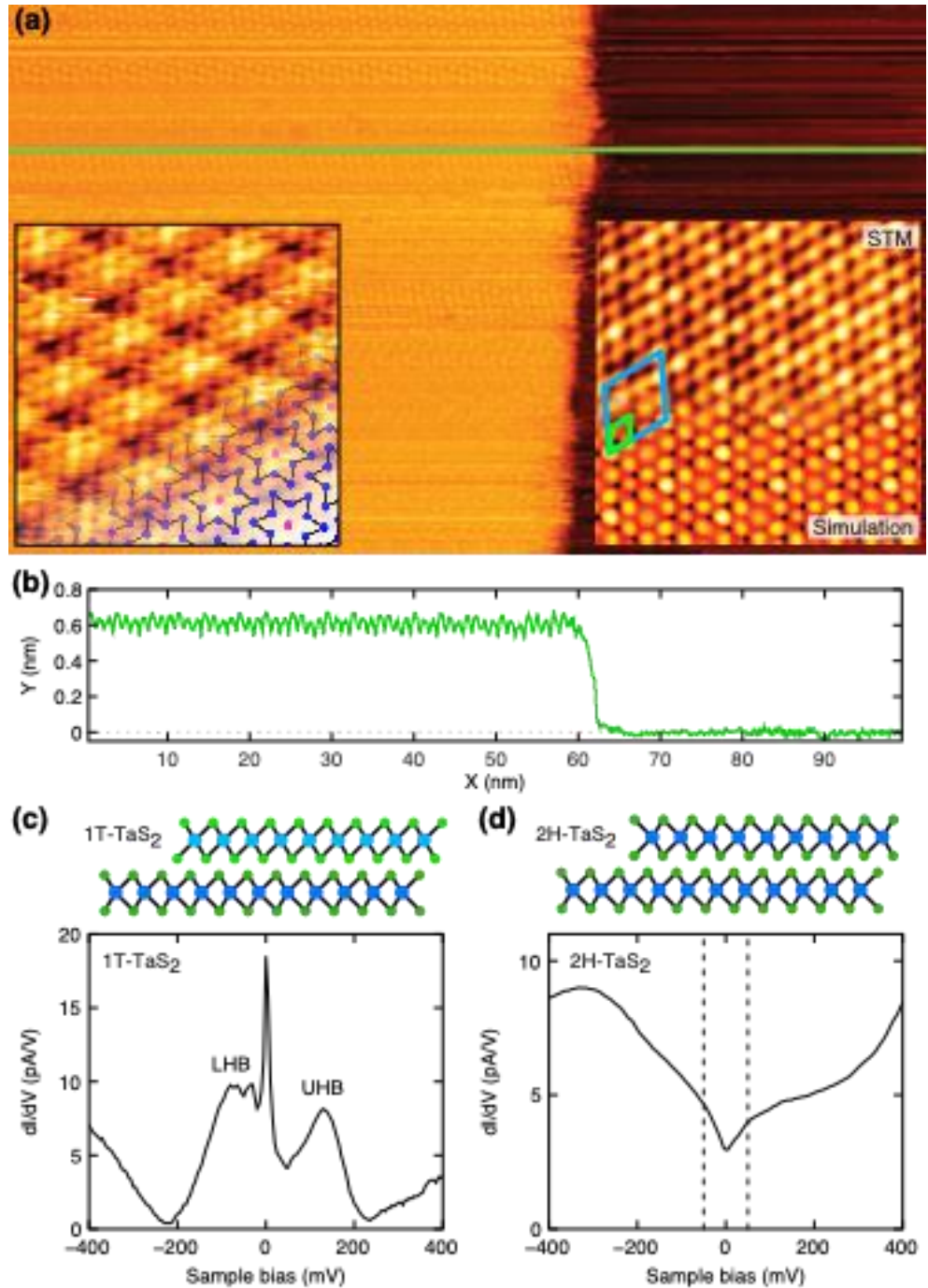

**Figure 2. (a)** Large scale STM topographic image showing two terraces on the 2H-TaS$_2$ surface separated by a single layer step-edge. Image parameters: $V_b$ = 500 mV, $I_t$ = 300 pA, size 100 nm x 60 nm. The terraces present different surface periodicities. The inset on the left-hand side shows an atomically resolved STM image acquired on the corresponding upper terrace. The $\sqrt{13} \times \sqrt{13}R13.9°$ CDW periodicity is clearly seen on the image, where a model of the SoD structure is superimposed. Image parameters: $V_b$ = 500 mV, $I_t$ = 300 pA, size 18 nm x 18 nm. The inset on the right-hand side shows an atomically resolved STM image of the quasi-(3×3) CDW present on the lower terrace combined with the simulated STM image resulting from the DFT calculations (see section 13 in the SI). The atomic periodicity is marked with the green rhombus and the quasi-(3×3) CDW with the blue rhombus. Image parameters: $V_b$ = -50 mV, $I_t$ = 200 pA, size 15 nm x 15 nm. **(b)** Line profile across the step edge shown in panel (a) with a solid green line, the single layer step-edge has a height of $619 \pm 50$ pm. **(c)** Schematic model of the stacking sequence on the upper terrace and the corresponding experimental STS spectrum. STS parameters: $V_b$ = 400 mV, $I_t$ = 500 pA, $V_{mod}$=4 mV. The upper and lower Hubbard sub-bands (marked UHB and LHB respectively) are separated by a sharp resonance appearing close to the Fermi level. **(d)** Schematic model of the lower terrace and corresponding STS spectrum. The pseudo gap produced by the quasi-(3×3) CDW is observed around zero bias (dashed vertical lines). STS parameters: $V_b$ = 400 mV, $I_t$ = 500 pA, $V_{mod}$=10 mV.

**Figure 2c** shows the STS spectroscopy data measured at 1.2 K on a terrace presenting the quasi-(3×3) CDW, with a sketch illustrating the corresponding stacking sequence. The expected width of the pseudo gap associated to the CDW[37] is marked with dashed vertical lines. Figure 2c shows an STS spectrum measured on the terrace presenting the $(\sqrt{13} \times \sqrt{13})R13.9°$ CDW, and the corresponding sketch of its stacking sequence. The Upper and Lower Hubbard sub-bands (marked as UHB and LHB) can be seen on both sides of the Fermi level. A third spectral feature is a sharp peak that appears close to the Fermi level. The STS spectra measured at 52 K on these 1T/2H areas still show the Hubbard sub-bands but the peak at the Fermi level is no longer present (see SI section 6). As it will be demonstrated in the following, we identify this peak as a Kondo resonance, resulting from the screening of the localized electrons at the center of the SoD clusters by the conduction electrons from the metallic substrate.[44]

**Figure 3a** shows the periodic spatial modulation in the intensity of the Kondo peak within the unit cell of the $(\sqrt{13} \times \sqrt{13})R13.9°$ CDW. The inset shows the corresponding STM topography, with green colored dots indicating the location where the representative STS spectra were measured. The Kondo resonance is more pronounced when the tip is located at the center of the SoD and its intensity drops rapidly when the tip moves slightly off center. In order to determine the Kondo temperature, we focus on the temperature evolution of the intrinsic full width at half maximum (FWHM) 2Γ. The Kondo resonance can be fitted with a Fano line-shape (equation (1)) to which we add some energy broadening to account for the experimental conditions, as explained in sections 7 and 8 of the SI.

$$\boldsymbol{\rho}(E) \propto \boldsymbol{\rho_0} + \frac{\left(q+\frac{E-E_0}{\Gamma}\right)^2}{1+\left(\frac{E-E_0}{\Gamma}\right)^2} \tag{1}$$

The three free parameters of the Fano line-shape are the intrinsic width $\Gamma$, the profile parameter q and the energy offset $E_0$. **Figure 3b** shows four selected spectra of the Kondo resonance from a temperature series acquired between 1.7 K and 20 K, see section 9 of the SI for the complete series. The black dots are the background-subtracted experimental data and the red curve is the fitting to the broadened Fano line shape. **Figure 3c** shows as orange squares the values of the fitted parameters for the complete series as a function of temperature. As expected, the q and $E_0$ parameters do not vary with temperature, while there is a clear evolution in $\Gamma$. According to the Fermi-liquid model,[29] the expected evolution of the intrinsic FWHM with temperature is given by the following expression:

$$2\Gamma = \sqrt{(\alpha k_B T)^2 + (2 k_B T_K)^2} \qquad (2)$$

where $T_K$ represents the Kondo temperature and $\alpha$ is a parameter that takes into account the thermal smearing during the Kondo process.[29] Figure 3c shows as a red line the fit of equation (2) to our experimental data, leaving $\alpha$ and $T_K$ as free parameters. The Kondo temperature resulting from this fit is 27 K, in agreement with the fact that no Kondo resonance is detected at 52 K (see SI section 6).

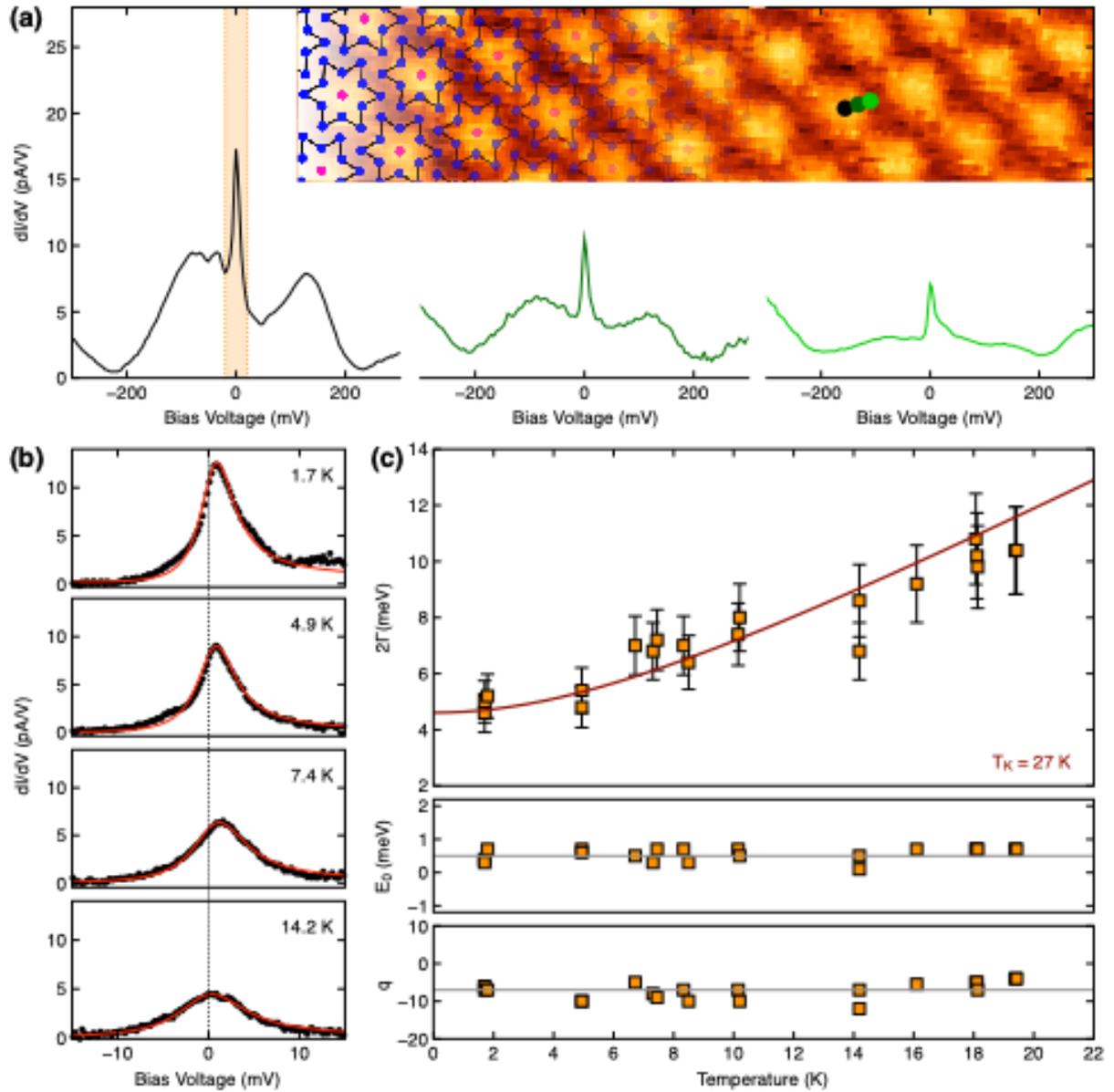

**Figure 3. (a)** Single point STS taken at 1.5K on different locations along the CDW unit cell, showing the modulation of the intensity of the zero bias Kondo peak. The Kondo resonance is more intense at the center of the SoD (black spectrum) and its intensity drops away from the center (dark and light green spectra). STS parameters: $V_b$ = 500 mV, $I_t$ = 500 pA, $V_{mod}$ = 4 mV. The inset shows the area of the sample where the spectra were measured. Image parameters: $V_b$ = 500 mV, $I_t$ = 90 pA, size 20 nm × 4 nm. **(b)** Selected STS spectra from a temperature series acquired at the center of the SoD. The black dots are the experimental data, the red curve is the fitting to a Fano line-shape. STS parameters $V_b$ = 25 mV, $I_t$ = 500 pA, $V_{mod}$ = 4 mV. **(c)** Temperature evolution of the Fano fitting parameters of the complete series. From this fit we obtain a Kondo temperature of 27K and an α value of 6.4.

When the 1T-TaS$_2$ single layer is placed on top of the 2H-TaS$_2$ substrate the system has all the key ingredients to develop a Kondo lattice. For temperatures below T$_K$ = 27 K, the unpaired electrons localized at the center of the SoD clusters are Kondo screened by the metallic 2H-TaS$_2$ substrate. At the same time the $(\sqrt{13} \times \sqrt{13})R13,9°$ CDW imposes a triangular 2D lattice with spatial long-range order that frustrates a possible antiferromagnetic arrangement of the magnetic moments located at the center of the SoD.[45] As discussed before, the formation of the quantum coherent state at low temperature, known as Kondo lattice, modifies the Fermi surface of the system in order to include in the periodic system the originally localized electrons. In **Figure 4a** we present a high-resolution STS spectrum measured on the heterostructure at 1.5 K using a bias modulation of 200 μV, which shows the appearance of a double peak structure that we attribute to the Kondo lattice gap, see Figure 1d. Note that this feature is not visible in the spectra measured with a larger (4 mV) bias modulation (see Figure 3 and section 9 in the SI). Also, due to the spatial modulation of the Kondo resonance, the gap feature is better resolved at the center of the SoD clusters, hence that's the selected location for all the STS measurements unless stated otherwise.

The formation of a Kondo lattice ensures the periodicity of the magnetic impurities and the creation of a flat band at the Fermi level. The hybridization of this band with the dispersing bands describing the conduction electrons of the metallic substrate is the origin of the heavy-fermion liquid phase. The main physics of the system can be described by the effective Hamiltonian:[23]

$$\mathbf{H} = \sum_{\mathbf{k}n} \epsilon_{\mathbf{k}n} d^\dagger_{\mathbf{k}n} d_{\mathbf{k}n} + \sum_{\mathbf{k}} \varepsilon_{\mathbf{k}i} c^\dagger_{\mathbf{k}i} c_{\mathbf{k}i} + \sum_{\mathbf{k}n} (\langle \mathbf{i}|\hat{V}_\mathbf{k}|\mathbf{n}\rangle c^\dagger_{\mathbf{k}ni} d_{\mathbf{k}n} + \text{h. c.}) \qquad (3)$$

where $\mathbf{k}$ is a reciprocal lattice vector inside the first Brillouin Zone, $d^\dagger_{kn}$ and $d_{kn}$ are fermionic operators for the conduction band $\epsilon_{kn}$ (with *n* a multi-index including multiple bands and spin) of the substrate; $c^\dagger_{ki}$ and $c_{ki}$ are the fermionic operators of the lattice of impurity states, with energies $\varepsilon_{ki}$, after they establish coherence (the "impurity band"); and $\hat{V}_k$ is the interaction potential, which, in general, depends on $\mathbf{k}$ and might be different for different conduction bands (see also SI section 12).

The 2H-TaS$_2$ crystals are metallic due to two doubly degenerate bands crossing the Fermi level that originate from the Ta 5d half-filled orbitals. The dispersion of these bands is slightly modified by the formation of a quasi-(3×3) CDW.[34,38] DFT calculations for the 2H phase of the bulk TaS$_2$ have been carried out within the Projector Augmented Wave (PAW) method as implemented in Vienna Ab initio Simulation Package (VASP). Due to the dispersive nature of the inter-plane interaction of the TaS$_2$ layers in the bulk, van der Waals effects have been included through the Grimme D3 correction to the Perdew-Burke-Ernzerhof (PBE) functional. A detailed analysis of the electronic properties of bulk 2H-TaS$_2$ based on DFT calculations is provided in sections 13 and 14 in the SI. As can be seen in Figure S11, these calculations reproduce fairly well the CDW of 2H-TaS$_2$. The corresponding bands have been used as conduction bands $\epsilon_{kn}$ in equation (3). The impurity band $\varepsilon_{ki}$, on the other hand, appears with the $(\sqrt{13} \times \sqrt{13})R13.9°$ periodicity of the SoD CDW in the 1T-TaS$_2$ overlayer, and is "made of" electrons from the 2H-TaS$_2$ substrate, which screen the magnetic impurities of the 1T overlayer. Since very little is known a priori about this band, we have assumed a band dispersion typical of TaS$_2$ with a total band width of the order of T$_K$ and a periodicity commensurate to the (3×3) 2H-TaS$_2$ lattice, which is justified by the narrow width of this band and allows keeping the computational cost at a manageable level. In **Figure 4b** and **c** we provide the band structure and the DOS in **Figure 4d**, as obtained by numerically diagonalizing the Hamiltonian in equation (3) with a (k-independent) $V = |\hat{V}_k| = 5$ meV interaction potential between the impurity band and *all* the conduction bands of the (3×3) 2H-TaS$_2$ surface. In Figure 4(c), the opening of gaps around the Fermi level resulting from the avoided band-crossing can be seen. The resulting DOS shown in Figure 4d consists of a double peak structure near the Fermi level as opposed to the single peak structure obtained when the interaction potential was switched-off (see Figure S9 in the SI).

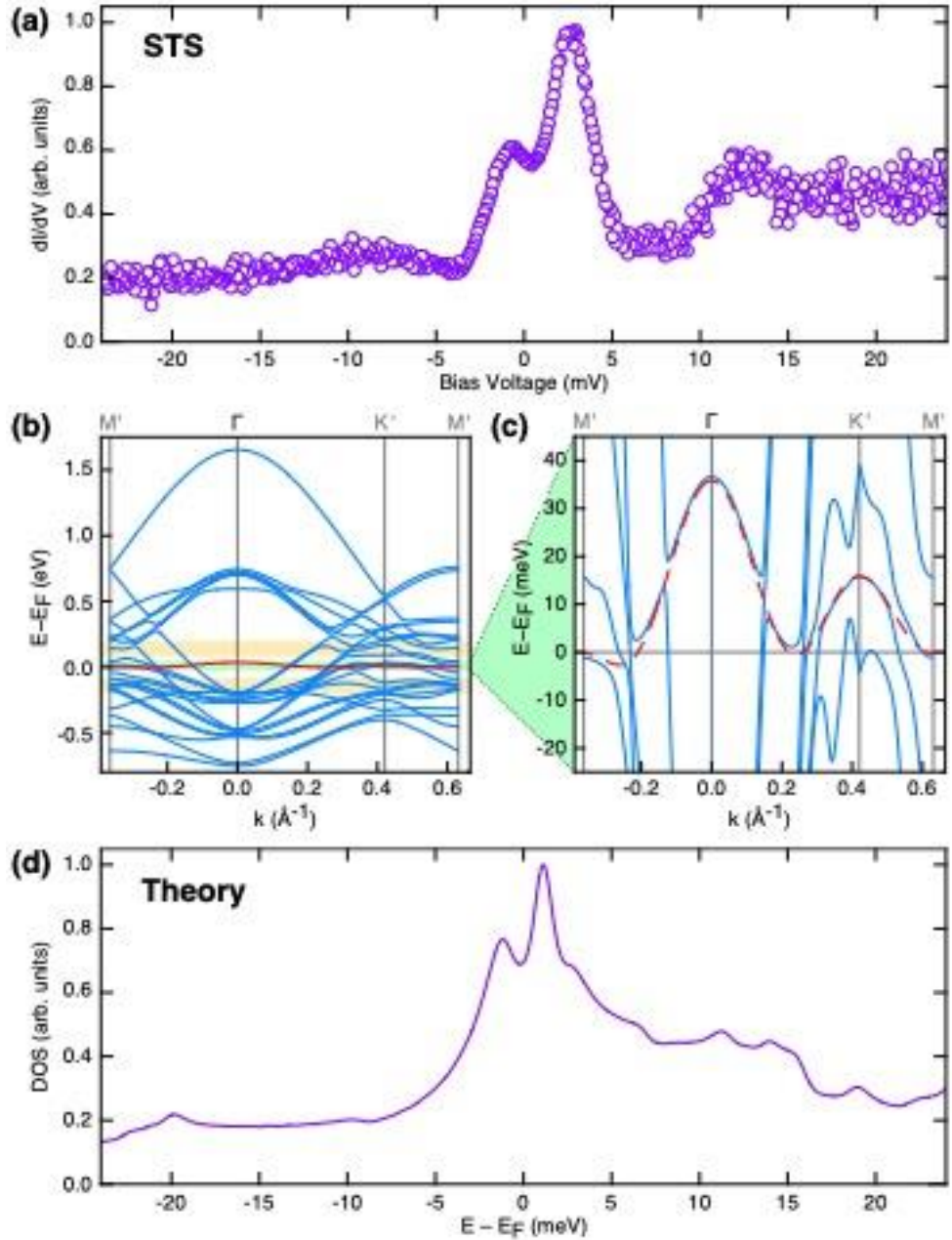

**Figure 4. (a)** High-resolution STS spectrum measured with 200 µV bias modulation at 1.5 K on the 1T/2H single layer region of the TaS$_2$ crystal. Two peaks separated by 4 meV are clearly resolved close to the Fermi level. **(b)** Blue lines show the band structure (along the high symmetry M'ΓK'M' BZ-path) obtained from the bands of the optimized (3×3) CDW reconstruction of the 2H-TaS$_2$ substrate for $k_z$=0. The impurity band is at the Fermi level (red line). The energy position of the Upper and Lower Hubbard sub-bands are marked with two yellow strips. **(c)** Zoom in of the band structure shown in panel (b) close to the Fermi level after an interaction potential $V_k = 5$ meV between the impurity band and all conduction states is switched-on (the unperturbed impurity band is depicted as a red-dashed line for reference). **(d)**

Calculated DOS for the system whose band structure is shown in panel (c): a two-peak structure at the Fermi level is predicted in agreement with the experiment.

## 3. Discussion

The Kondo lattice regime shows up at temperatures between 0 and $T_{KL}$, where $T_{KL}$ is the Kondo lattice temperature, which must be lower than $T_K$ as already explained. Therefore, the Kondo lattice gap should disappear at $T > T_{KL}$, leaving a "standard" Fano-line-shaped Kondo resonance for $T_{KL} < T < T_K$. **Figure 5a** shows four selected spectra from a temperature series recorded using 500 µV bias modulation (see section 10 in the SI for the full series), where the black dots represent the experimental data. All spectra have been measured at the center of the SoD in the 1T-TaS$_2$ layer lying on top of the 2H-TaS$_2$ crystal. The choice of a 500 µV bias modulation was a compromise between being able to resolve the Kondo lattice gap (for which a low bias modulation is needed) and minimizing the effect of thermal drift keeping a low acquisition time (which can be achieved increasing the bias modulation to improve the signal to noise ratio). The spectrum of Figure 5a measured at 1.7 K presents a structure compatible with the double peak shape observed in the spectra measured with a 200 µV bias modulation. The red curves in Figure 5a are fits to a broadened Fano line-shape excluding the data in the energy interval of ±3 meV around the Fermi level, as explained in detail in section 8 of the SI. In this way, we avoid the energy range where the opening of the Kondo-lattice gap takes place. For the spectra measured at 1.7 K, 4.2 K and 7.5 K there is a clear deviation between the experimental data and the fits, as highlighted by the blue shaded area. On the contrary, in the spectrum measured at 12.1 K the agreement between the experimental points and the Fano line shape is excellent. As expected, the overall intensity of the signal decreases with temperature.

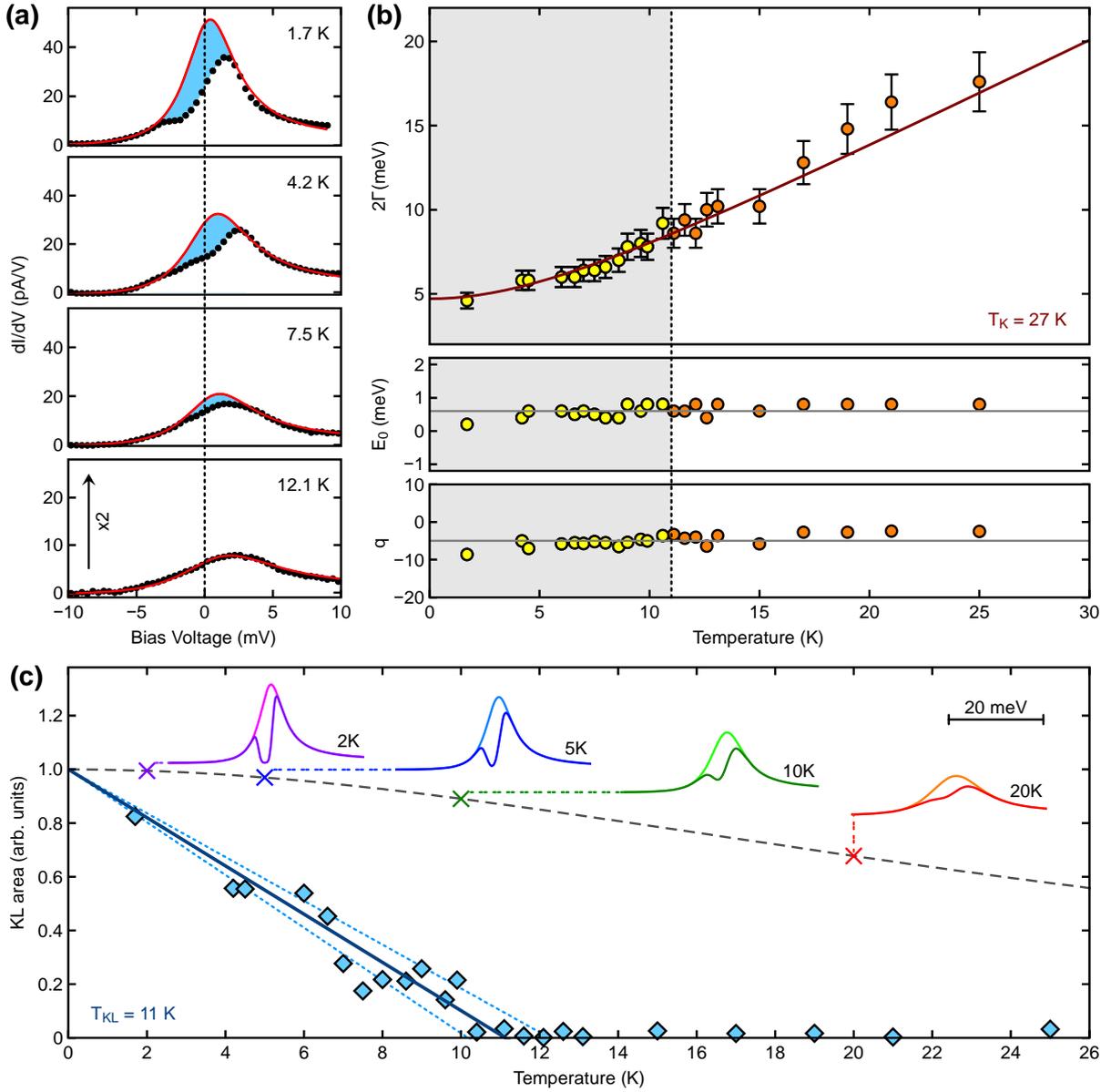

**Figure 5.** **(a)** STS spectra from a high energy resolution temperature series (black dots). The red lines are Fano fits excluding the ±3 mV energy range, as explained in the text. The blue shaded area highlights the difference between the "ideal" Fano line-shape and the actual data and represents the LDOS depletion associated to the Kondo lattice gap. Due to the intensity reduction of the Kondo signal at higher temperatures, the vertical scale for the spectrum measured at 12.1 K is upscaled by a factor 2. STS parameters: $V_b$ = 25 mV, $I_t$ = 500 pA, $V_{mod}$=500 $\mu$V. **(b)** Temperature evolution of the Fano fitting parameters of the complete series, represented as circles. Yellow instead of orange indicates a deviation from the ideal Fano fit. In the top graph, the temperature evolution of the intrinsic width at half maximum is fitted with eq. 2 (red curve) yielding a Kondo temperature of $T_K$ = 27 K and an $\alpha$ value of 7.6. The lower two panels show no trend in the profile parameter ($q$) or the energy position ($E_0$), as expected.

In addition, they are almost identical to the ones obtained for the low-resolution series of Figure 3. **(c)** Temperature evolution of the area associated to the Kondo lattice gap, represented by the blue shaded area in panel (a) and marked here as blue filled diamonds. The area becomes practically negligible above 10 K. The dotted blue lines show the range of possible linear fits based on the last experimental point included. This yields a Kondo lattice transition temperature $T_{KL} = 11K \pm 1$ K. The graphs inserted in the upper part of the plot in Figure 5c are the gapless (light color curves) versus gapped (dark color curves) Kondo resonances calculated for the ideal gap case for four selected temperatures, corresponding to the colored crosses displayed in the main plot. The gray dashed line represents the thermal attenuation effect on an ideal and temperature-independent 3 meV wide gap on a Kondo resonance with similar parameters as the one observed experimentally. The temperature dependence shows a much slower decay as compared to the observed Kondo lattice gap. In both cases the vertical scale is normalized so that the area at zero temperature is defined as 1.

**Figure 5b** shows the obtained fitting parameters for the complete temperature series as circles, where yellow or orange fill color indicates whether there was a deviation from the Fano line-shape or not, as explained in the previous paragraph. The upper graph displays the temperature evolution of the intrinsic width of the Fano line-shape together with a fit to equation (2), represented by a red line. This fit gives a Kondo temperature of 27 K, in good agreement with the value obtained from the spectra measured with lower resolution (see Figure 3). Note that both fits give slightly different α values (6.4 vs 7.6) that we attribute to slight differences in the temperature stabilization procedure employed in each case (see section 8 in the SI for detailed discussion) The lower graphs in Figure 5b show the temperature evolution of the other two parameters. Neither the profile parameter ($q$) nor the energy position of the Kondo resonance ($E_0$) change with temperature, reproducing the behaviour of the low-resolution series in Figure 3, as expected.

To analyze the suppression of the Kondo lattice features in our STS data as a function of temperature, we have computed the area difference between the "ideal" or "gapless" Fano line-shapes (red curves in Figure 5a) and the actual STS data (black dots), see section 8 of the SI for the details in the area computation. This area, which can be interpreted as the LDOS depletion associated to the Kondo lattice gap,[20,21] is highlighted in blue in Figure 5a and represented as a function of temperature in **Figure 5c** as blue filled diamonds, where the vertical scale is normalized so that the projected area at zero temperature is defined as 1. The area monotonically decreases as the temperature increases and vanishes above 11 K. To estimate the effect of thermal broadening on the experimentally observed Kondo lattice gap width we calculate the thermal attenuation of an ideal and temperature-independent 3 meV wide gap on a Kondo resonance modelled with the parameters obtained experimentally in this work (see SI section 11 for the details). The upper insets in Figure 5c illustrate the resulting area difference between these ideal gapless (light color) and gapped (dark color) Kondo resonances for four selected temperatures. The complete temperature evolution of the resulting gap area is represented by the gray dashed curve in the main plot, demonstrating that even at temperatures as high as 20K a deviation between the ideal Kondo resonance and the gapped one is clearly visible. Therefore, we conclude that the experimentally observed area drop presented in Figure 5c cannot be attributed to the thermal broadening of the gap. The observed gap reduction with temperature corresponds to the annihilation of the Kondo lattice gap due to the loss of coherence of the Kondo clouds, signaling the transition between the Kondo lattice regime and the single Kondo impurity regime. A linear fit to the experimental data, shown as a blue line in Figure 5c, yields a transition temperature of $T_{KL} = 11K \pm 1$ K. A similar linear dependence of the Kondo lattice gap with temperature has been observed in a YbRhSi crystal [46].

In summary, we have studied the temperature evolution of the electronic properties of a system composed of a single layer Mott insulator on top of a metal. We have found that, as one lowers the temperature down to 27K, the magnetic moments present in the Mott insulator start to experience the Kondo screening by the conduction electrons of the metal, leading to the appearance of a Kondo resonance at the Fermi level. By further lowering the temperature down to 11 K, the strength of the Kondo screening increases, as demonstrated by the increase in the intensity of the Kondo resonance. Below 11 K, a gap opens within the Kondo resonance, which is the signature of the formation of a coherent quantum state that extends all over the sample, i.e., a Kondo lattice. This state results from the overlap between the Kondo clouds associated with the local magnetic moments in the Mott insulator layer.

**Supporting Information**

Supporting Information is available from the Wiley Online Library.

**Acknowledgements**

This work was supported by Ministerio de Ciencia, Innovación y Universidades through grants, PID2021-128011NB-I00 and PID2019-105458RB-I00. Ministerio de Ciencia e Innovación and Comunidad de Madrid through grants "Materiales Disruptivos Bidimensionales (2D)" (MAD2D-CM)-UAM and "Materiales Disruptivos Bidimensionales (2D)" (MAD2D-CM)-IMDEA-NC funded by the Recovery, Transformation and Resilience Plan, and by NextGenerationEU from the European Union. Comunidad de Madrid through grants NMAT2D-CM P20128/NMT-4511 and NanoMagCost. IMDEA Nanoscience acknowledges support from the ''Severo Ochoa'' Programme for Centres of Excellence in R&D CEX2020-001039-S. IFIMAC acknowledges support from the ''María de Maeztu'' Programme for Units of Excellence in R&D CEX2018-000805-M. MG thanks Ministerio de Ciencia, Innovación y Universidades "Ramón y Cajal" Fellowship RYC2020-029317-I. Allocation of computing time at the Centro de Computación Científica at the Universidad Autónoma de Madrid, the CINECA Consortium INF16_npqcd Project and Newton HPCC Computing Facility at the University of Calabria (MP).